\documentclass[twocolumn]{aastex63}
\usepackage{amsmath,amstext}
\usepackage[T1]{fontenc}
\usepackage[figure,figure*]{hypcap}
\usepackage{multirow}
\usepackage[T1]{fontenc}
\usepackage[utf8]{inputenc}
\usepackage{float}
\newcommand{\ba}{\begin{eqnarray}}
\newcommand{\ea}{\end{eqnarray}}
\usepackage{lineno}

\def\b{_{\rm b}}
\def\c{_{\rm c}}

\newcommand{\jhat}{\hat{\textbf{\j}}}

\def\i{_{\rm bc}}

\def\p{_{\rm p}}
\def\d{_{\rm d}}

\def\jin{{\bf j}_{\rm bc}}

\newcommand{\x}{\sout}
\newcommand{\y}{\color{red}}
\newcommand{\w}{\color{blue}}

\graphicspath{{./}{figures/}}

 \received{November 2021}
 \accepted{January 2022}

\shorttitle{A chaotic origin for K2-290}
\shortauthors{Best \& Petrovich}

\begin{document}

\title{The chaotic history of the retrograde multi-planet system in K2-290A driven by distant stars}

\correspondingauthor{Marcela Best}
\email{mbest@uc.cl}

\author[0000-0001-8361-9463]{Marcela Best}
\affiliation{Instituto de Astrofísica, Pontificia Universidad Católica de Chile, Av. Vicuña Mackenna 4860, 782-0436 Macul, Santiago, Chile}
\author[0000-0003-0412-9314]{Cristobal Petrovich}
\affiliation{Instituto de Astrofísica, Pontificia Universidad Católica de Chile, Av. Vicuña Mackenna 4860, 782-0436 Macul, Santiago, Chile}
\affiliation{Millennium Institute for Astrophysics, Chile}

\begin{abstract}
The equator of star K2-290A was recently found to be inclined by $124\pm 6$ degrees relative to the orbits of both its known transiting planets. The presence of a companion star B at $\sim 100$ au suggested that the birth protoplanetary disk could have tilted,
thus providing an explanation for the peculiar retrograde state of this multi-planet system.
In this work, we show that a primordial misalignment is not required and that the observed retrograde state is a natural consequence of the chaotic stellar obliquity evolution driven by a wider-orbit companion C at $\gtrsim2000$ au long after the disk disperses.
The star C drives eccentricity and/or inclination oscillations on the inner binary orbit, leading to widespread chaos from the periodic resonance passages between the stellar spin and planetary secular modes. Based on a population synthesis study, we find that the observed stellar obliquity is reached in $\sim 40-70\%$ of the systems, making this mechanism a robust outcome of the secular dynamics, regardless of the spin-down history of the central star.
This work highlights the unusual role that very distant companions can have on the orbits of close-in planets and the host star's spin evolution, connecting four orders of magnitude in distance scale over billions of orbits. We finally comment on the application to other exoplanet systems, including multi-planet systems in wide binaries.

\vspace{1cm}

\end{abstract}

\section{Introduction}

\begin{figure*}[]\center
\includegraphics[width=12cm]{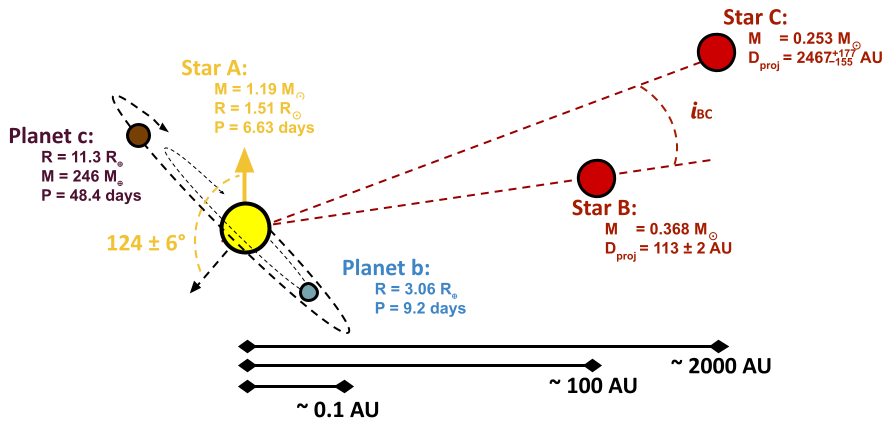}
\caption{Schematic depiction of the K2-290 system. The planets b and c are nearly coplanar with an orbital plane inclined by $\sim 124^\circ$ relative to the equator of star A. The distant M-dwarfs B and C have projected distances of $\sim100$ au and $\sim 2000$ au, respectively, and we define their mutual inclination as $i_{\rm BC}$.}
\label{fig:schematic}
\end{figure*}

The K2-290A star (EPIC 249624646; an F8 star of 1.19 M$_\odot$) was found to host two transiting planets: a mini-Neptune at a $9-$day orbit and warm Jupiter with an orbital period of 48 days \citep{Hjorth_2019}. From adaptive optics imaging and Gaia data the authors identify two M-dwarf companions at projected separations of $\sim 110$ au and $\sim 2500$ au. You can find all the relevant parameters in Figure \ref{fig:schematic}.

Furthermore, not only is K2-290 a triple stellar system hosting planets, but it was later found in \citet{Hjorth_2021} that both of its planets have retrograde stellar obliquities\footnote{The stellar obliquity of planet c is well determined, while that of b is less so, but still firmly retrograde. Since $b$ and $c$ are likely nearly coplanar, a single absolute obliquity of $124\pm 6$ degrees can be determined for both planets b and c.} ($124\pm 6$ degrees obliquity with respect to their host star spin axis). This retrograde state is unique among the current sample of compact multi-planet systems with nearly a dozen well-aligned ones and a handful of misaligned, still prograde, ones (see the recent discussion in \citealt{wang2021}).

The extreme orbital tilt and the unusual presence of two companions stars suggest a peculiar dynamical evolution for K2-290. Other systems with misaligned ($\psi\sim 40^\circ$) multi-planet systems include Kepler-56 \citep{kepler56,kepler56d} and Kepler-129 \citep{kepler129}, and may be explained by interactions with the detected distant Jovian-sized planets in the systems \citep{BF2014,li2014,GF2017}. A more extreme example is the polar system HD3167 \citep{HD3167}, which may still be accommodated by interplanetary interactions by a yet undetected companion \citep{BF2014,petro2020}.

Although retrograde systems do exist, these correspond to lonely short-period planets---hot Jupiters and warm Netpunes. These systems likely acquired their stellar obliquities via a different path, possibly by high-eccentricity migration (e.g., \citealt{dawson_review}), a process that is forbidden for multi-planet systems like K2-290 \citep{mustill2015}.

An explanation provided by \citet{Hjorth_2021} for the retrograde orbits of K2-290b and c is that the birth protoplanetary disk tilted relative to the host star's equator \citep{batygin12}. Here, an inclined binary companion $B$ would tilt the disk, while the disk dispersal leads to a secular resonance that tilts the host star spin's axis sometimes reaching retrograde states \citep{lai2014,Spalding2014}. Although this may be a reasonable possibility and calculations show that it can reproduce the system's obliquity for typical parameters, there is a fair amount of uncertainty, including the disk size and dispersal history, planet migration, star-disk coupling, and so on, that may inhibit the occurrence of the resonance \citep{jj2018}. Moreover, the questions of whether the stellar obliquity in the disk phase becomes the observed obliquity after billions of years and the role of the third star C remain unaddressed.

Motivated by these questions and by the striking fact that all the relevant precession timescales in the system are comparable, we study the long-term evolution of K2-290's stellar obliquity. Our paper is organized as follows. In \S\ref{procedure} we describe our methods. In \S\ref{results} we show our results, a population synthesis and a physical description of the mechanism that drives obliquity excitation. We discuss our results in \S\ref{discussion} and present our conclusions in \S\ref{conclusions}.





\section{Procedure}\label{procedure}

We simulate the long-term evolution of the system using a secular set of equations described in Appendix \ref{app:equations_of_motion}. This allows us to quickly evolve the system for up to 5.6 Gyrs (the upper limit on the age estimate for the central star).

As some parameters are not well constrained (e.g. the mass of planet b, distance to the companion stars, initial rotational period of the central star, etc) we explored this parameter space and see which combinations reproduce the desired outcome.

\subsection{Simplifying the problem}


The K2-290 system is composed of fives bodies (three stars, two planets), which would mean 10 pair-wise interaction terms. However, only a fraction dominates the dynamics and this allows us to simplify the problem.

The most important simplification is to consider the two planets effectively locked together (i.e. sharing their orbital planes). This coupling is justified using equation 12 from \citet{Lai_2016}:
\begin{equation}
\epsilon_{\rm b,c} \approx \left( \frac{M_{\rm B}}{10^3m_{\rm c}} \right) \left( \frac{10a_{\rm c}}{a_{\rm B}} \right)^3 \left[ \dfrac{3a_{\rm b}/a_{\rm c}}{b^{(1)}_{3/2}(a_{\rm b}/a_{\rm c})} \right] \dfrac{(a_{\rm c}/a_{\rm b})^{3/2} - 1}{1 + (L_{\rm b}/L_{\rm c})},
\end{equation}
which results in $\epsilon_{\rm b,c}\sim 2 \times 10^{-5}$ for the values in Figure \ref{fig:schematic}, assuming M$_{\rm b}$ = 7.6 M$_{\Earth}$, this means that the two planets are strongly coupled together and their mutual inclination could only reach values in the same order when perturbed by star B. A similar conclusion is reached comparing with stellar $J_2$. Also, we assume their orbits remain circular \footnote{In \citet{Hjorth_2019} there is the possibility for a nonzero eccentricity of planet c, although reported values are for circular orbits and to 3$\sigma$ they report an upper limit on the eccentricity of 0.241.} as the precession rate for planet c due to its inner companion is higher than the precession induced by star B even for high values of $e_B \sim 0.99$ (equation 10 from \citealt{naoz2018}).

Further simplifications include planets b and c interacting with star C only indirectly via star B, justified since the torque from C is weaker than that of B by a factor of $\left(b_{\rm B}/b_{\rm C}\right)^3\sim 10^{-4}$. We also assume that the planets do not torque stars B and C.

\subsection{Precession frequencies}

From the equations of motion in Appendix \ref{app:equations_of_motion}, we can evolve the stellar spin axis and planetary angular momentum vectors as
\ba
\frac{d{\bf s}}{dt}&=& \dot{\Omega}_{\rm s,bc} ~\left({\bf s}\times \hat{{\bf j}}_{\rm bc}\right), \\
\frac{d\hat{\bf j}_{\rm bc}}{dt}&=& \dot{\Omega}_{\rm bc,s} ~\left(\hat{\bf j}_{\rm bc}\times {\bf s}\right)+
\dot{\Omega}_{\rm bc,B} ~\left(\hat{\bf j}_{\rm bc} \times {\bf j}_{\rm B} \right),
\ea
where $\hat{\bf{j}}_{\rm bc}$ is the unit vector of the angular momentum of both planets and $\dot{\Omega}$ are precession frequencies. The magnitudes of angular momenta are $L_{\rm bc}=m_{\rm b}\sqrt{\mathcal{G}M_{\rm A} a_{\rm b}}+m_{\rm c}\sqrt{\mathcal{G}M_{\rm A} a_{\rm c}}$ and that of the spin is $S = k_q M_{\rm A} R_{\rm A}^2 \Omega_{\star}$.

The precession frequencies are approximated by
\ba
\label{eq:Omega_bc_s}
\dot{\Omega}_{\rm bc,s}&=&
\frac{3J_2(GM_{\rm A})^{1/2} R_{\rm A}^2}{2 (m_{\rm b}\sqrt{ a_{\rm b}}+m_{\rm c}\sqrt{a_{\rm c}})} \left( \frac{m_{\rm b}}{a_{\rm b}^3} + \frac{m_{\rm c}}{a_{\rm c}^3} \right) \cos(i_{\rm s,bc}) \nonumber\\
&\simeq &0.28 \mbox{Myr}^{-1} \left[ 0.5 + 0.45 \left( \frac{33 m_{\rm b}}{m_{\rm c}} \right) \right] \nonumber\\
& &\times\left( \frac{J_2}{3\times 10^{-6}} \right)\cos(i_{\rm s,bc})
\ea
where we have used the quadrupole moment of the central star $J_2 \approx \frac{1}{3} k_2 \left( \Omega_{\star}^2 R_{\rm A}^3 / GM_{\rm A} \right) $ \citep{war76} with $k_2$ its stellar Love number and
\ba
\label{eq:Omega_s_bc}
\dot{\Omega}_{\rm s,bc}=
\dot{\Omega}_{\rm bc,s}\left(\frac{S}{L}\right)\simeq0.45\left(\frac{6.63\mbox{ d}}{P_{\star}}\right)
\dot{\Omega}_{\rm bc,s},
\ea
where $P_\star$ is the period of the central star and we have assumed that $k_q=0.06$.

In turn, for the planets and star B we get:
\ba
\label{eq:Omega_bc_B}
\dot{\Omega}_{\rm bc,B}&=&\frac{3G^{1/2} M_{\rm B} a_{\rm c}^{3/2}}{4a^3_{\rm B} M_{\rm A}^{1/2}} \frac{\cos(i_{\rm bc,B})}{(1-e_{\rm B}^2)^{3/2}}\nonumber \\
&\simeq& 0.23 \mbox{Myr}^{-1} ~\left(\frac{100\mbox{ au}}{a_{\rm B}}\right)^3 \frac{\cos(i_{\rm bc,B})}{(1-e_{\rm B}^2)^{3/2}}, \label{bc,B}
\ea
where we have assumed that the star B only couples to the outer planet c. This is reasonable as the outer planet will precess faster due to B by a factor of $(a_{\rm c}/a_{\rm b})^{3/2}=5.25$ compared to planet b.

Finally, the A-B binary undergoes nodal precession and/or eccentricity/inclination oscillations due to star C in a von Zeipel-Lidov-Kozai (ZLK) oscillations
timescale \citep{vonzeipel,kozai62,lidov62}:
\ba
\tau_{\rm ZKL}&=&\frac{2P_{\rm B}}{3 \pi}\left(\frac{b_{\rm C}}{a_{\rm B}}\right)^3\frac{(M_{\rm A}+M_{\rm B})}{M_{\rm C}}\nonumber\\
\simeq 1.55 &\mbox{ Myr }& \left(\frac{a_{\rm B}}{100\mbox{ au}}\right)^{3/2}\left(\frac{b_{\rm C}}{20 a_{\rm B}}\right)^3\frac{(M_{\rm A}+M_{\rm B})}{M_{\rm C}}.
\ea

From these estimates we note a crucial result from this paper that {\it all timescales are in the same order of magnitude} for K2-290. Secular commensurabilities can occur for reasonable values of the system and when they do not, we will show that ZKL cycles can modulate $e_{\rm B}$ so that $\dot{\Omega}_{\rm bc,B} \propto 1/(1-e_{\rm B}^2)^{3/2}$ periodically \footnote{Strictly speaking, $\dot{\Omega}_{\rm bc,B}$ can cross the other frequencies at points other than the ZKL eccentricity peaks, but these cause the most dramatic changes to the system.} crosses $\dot{\Omega}_{\rm bc,s}$ and/or $\dot{\Omega}_{\rm s,bc}$.




\subsection{Spin down}

Even though the current value of the period for star A is known to be 6.63 days, presumably the star started with a shorter period ($P_{\star,0}$) and then, due to magnetic braking it slowed down \footnote{Although, given its short period, this effect was probably not significant for star A.} (e.g. \citealt{Anderson_2018}). To study the sensitivity of our results on the initial period of the star we assume the Skumanich law for the spin-down \citep{skumanich,bouvier}:
\begin{equation}
    \Omega_\star(t) = \dfrac{\Omega_{\star, 0}}{\sqrt{1+\alpha_{\rm MB} \Omega_{\star, 0}^2 t}},
\end{equation}
where $\Omega_{\star,0}$ is the initial angular velocity of star A and the constant $\alpha_{\rm MB}$ was chosen so that the period at the current age of the star (about 4 Gyr) coincides with the current period of star A.

\subsection{Planet b mass}

\citet{Hjorth_2019} give a mass estimation of 7.6 M$_{\Earth}$ using the mass-radius relation from \citet{Weiss_2014}. It is worth noting that in \citet{Hjorth_2019} this mass could not be observationally constrained and was only found to be 5.8 $\pm$ 5.1 M$_{\Earth}$. Therefore, we shall vary m$_{\rm b}$ in our population synthesis in a wide range of 1-15 M$_{\Earth}$.

Using the relation from \citet{kipping} we get a mass of 9.8 M$_{\Earth}$ which is close to the previously cited value and within the observational constrain. As it will later be shown in Figure \ref{fig:fig4}d, our results do not depend much on this parameter and so we adopt the mass from \citet{Hjorth_2019} as our fiducial value.

\begin{figure}[h!]
\includegraphics[width=9cm]{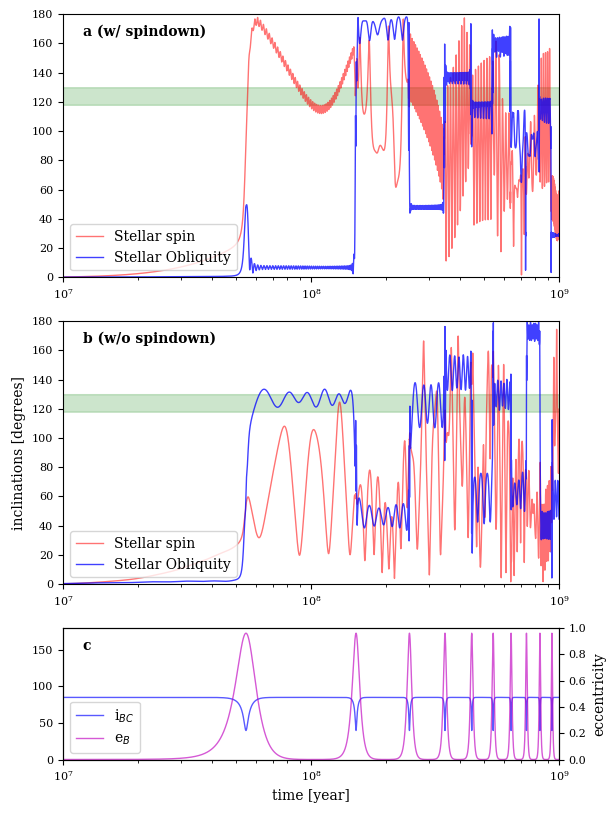}
\caption{Evolution of the inclination of the stellar spin axis ($i_{\rm s}=\cos^{-1}[\hat{\bf s}\cdot \hat{\bf z}]$) and the stellar obliquity ($i_{\rm bc, s}=\cos^{-1}[\hat{\bf s}\cdot \hat{\bf j}_{\rm bc}]$) for two examples using the fiducial values given in Figure \ref{fig:schematic} and final $J_2=10^{-5.6}$. \textbf{Panel a} shows the case with spin-down of the host star from $P_0=2$ days to its current value of 6.63 days, while \textbf{panel b} shows the same case without spin-down. 
In \textbf{panel c}, we show the ZKL cycles of the AB binary and observe that sudden and chaotic changes in the obliquities coincide with the instances when the eccentricity peaks are reached.}
\label{fig:fig1}
\end{figure}

\section{Results} \label{results}

We show two examples of successful simulations (i.e. achieving obliquities of $124^\circ$) in Figure \ref{fig:fig1} for our fiducial parameters including the
von Zeipel-Lidov-Kozai (ZLK) oscillations
\citep{vonzeipel,kozai62,lidov62} to emphasise that this is what kicks the system out of the equatorial plane and decouples the inner planets from their star in most cases.

\begin{figure}[h!]
\includegraphics[width=9cm]{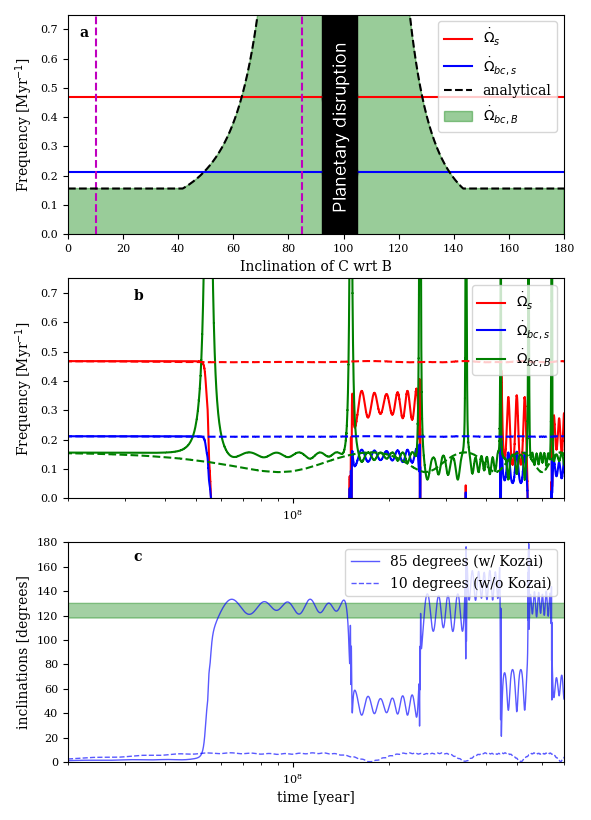}
\caption{Nodal precession frequencies $\dot{\Omega}_{\rm s,bc}$ (red lines), $\dot{\Omega}_{\rm bc,s}$ (blue lines) and $\dot{\Omega}_{\rm bc, B}$ (green lines and shaded region) for the example in Figure \ref{fig:fig1}b comparing two cases: mutual inclination $i_{\rm B,C}=10^\circ$ (no ZKL cycles; dashed lines) and $i_{\rm B,C}=85^\circ$ (ZKL cycles, solid lines). {\bf Panel a:} dependence on the initial mutual inclination $i_{\rm B,C}$ with the shaded green region displaying the whole possible range for $\dot{\Omega}_{\rm bc, B}\propto \left[1-e_{\rm B}^2\right]^{3/2}$; resonances are only encountered for $i_{\rm B,C}\gtrsim50^\circ$.
{\bf Panel b:} time evolution showing the resonance crossings that occur only for initial $i_{\rm B,C}=85^\circ$ (solid lines). {\bf Panel c:} evolution of the stellar obliquity showing that the sudden changes coincide with resonance crossings. The case with initial $i_{\rm B,C}=10^\circ$ without resonance crossing have only small periodic changes.
}
\label{fig:fig2}
\end{figure}

\begin{figure*}[t!]
\includegraphics[width=18cm]{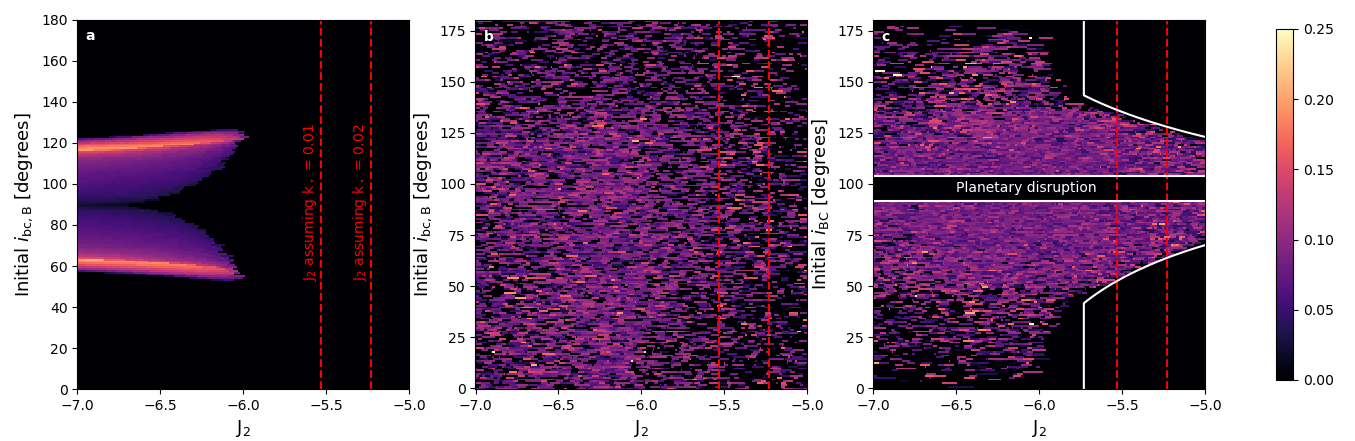}
\caption{Fraction of the time that the system spends with an inclination of 124$\pm$ 6 within the system's age (3.2-5.6 Gyr). Parameters where set according to Figure \ref{fig:schematic}, while varying the stellar inclinations (vertical axes) and $J_2$ in a broad range. The red vertical lines indicate the current values of $J_2$ for Love numbers $k_2$ of 0.01 and 0.02. {\bf Panel a:} Case without star C. {\bf Panels b and c:} case with star C with fractions as a function of the initial values of $i_{\rm bc, B}$ and $i_{\rm B,C}$, respectively. The white lines correspond to the limiting condition to cross the resonances in Equation (\ref{eq:J2_cosI}) and the horizontal ones correspond to the limit in which the system is destroyed because star B approaches too much the inner planets {\bf ($a_B[1-e_B]\approx a_c$)}.}
\label{fig:fig3}
\end{figure*}

\subsection{Secular chaos: conditions and available phase-space}

When star C is sufficiently inclined relative to AB, the eccentricity of star B (e$_{\rm B}$) reaches high values via ZKL oscillations which modulate the precession frequency of the inner planets as peaks in eccentricity translate into peaks in the nodal precession rate ($\dot{\Omega}_{\rm bc,B} \propto 1/(1-e_{\rm B}^2)^{3/2}$). This can be seen in Figure \ref{fig:fig2} which compares a case of low mutual inclination (i$_{\rm BC}$) to a higher one reaching higher eccentricities. This modulation of the potential can knock over the orbit of the planets out of the equatorial plane or decouple them from their star. We can calculate an analytical maximum value of this precession rate by knowing the maximum eccentricity of the cycles and using equation \eqref{bc,B}. It is worth noting that the angular momentum of stars B and C are comparable, which is why the effect does not peak at 90 degrees as it would for a test particle system.

Also e$_{\rm C}$ plays a role here but is assumed zero to compare with the analytical equation given in \citet{Naoz_2013} (we later consider the case of e$_{\rm C} \neq 0$ for the population synthesis):

\ba \label{naoz}
\gamma^2e_{\rm B, max}^4+\left[ 3+4\gamma \cos i_{\rm BC}+ \frac{\gamma^2}{4} \right] e_{\rm B, max}^2
\ea
\vspace{-0.5cm}
\ba
+ \gamma \cos i_{\rm BC} - 3 + 5 \cos ^2i_{\rm BC} = 0
\nonumber
\ea
Where $\gamma = L_{\rm B}/L_{\rm C}$. We can see in Figure \ref{fig:fig2}b that for an inclination of 10 degrees, $\Omega_{\rm bc,B}$ does not cross the other precession rates, but when we move to 85 degrees and start the ZKL cycles they do intersect which then knocks the system over.

In Figure \ref{fig:fig3} we run a simplified population synthesis fixing various parameters ($a_B$, $a_C$, $m_b$, $e_{\rm B}=e_{\rm C}=0$, no spin-down) to gauge the role of the star C.
 We can see that when star C is added (panels b and c), there is a wider parameter space in which we get to the desired obliquity, which appears to be fairly isotropic with i$_{\rm sB}$, compared to the case without star C in which we need values of i$_{\rm bc,B}$ close to 85 degrees. It is also worth noting that for our estimation of the value of J$_2$, using a Love number of 0.01 and 0.02
\footnote{As star A should be entering the sub-giant branch and so its convective zone should be deepening, this range in $k_2$ covers all the way from fully radiative to fully convective stars \citep{batygin2013}). }
and using the star parameters given in Figure \ref{fig:schematic}, we cannot get the retrograde obliquity without star C, at least when $e_B=0$.

In panel c of Figure \ref{fig:fig3} we observe empty regions marked by white lines where extreme obliquities are never reached. First, the horizontal band around $i_{\rm BC}\sim95^\circ$ is empty simply because the star B achieves eccentricities close to 0.99 and the system is destroyed as star B approaches too much to the planets {\bf ($a_B(1-e_B)\approx a_c$)}.

Second, the envelopes at large $J_2$ indicate the empty regions where the precession rate of the planets due to star A's $J_2$ is greater than the one due to star B ($\dot{\Omega}_{\rm bc,s} >\dot{\Omega}_{\rm bc,B}$). Equating the precession rates from Equations (\ref{eq:Omega_bc_s}) and (\ref{eq:Omega_bc_B}) at the maximum eccentricity $e_{\rm B, max}$, we get the following condition
\ba
\label{eq:J2_cosI}
\frac{J_2}{1.9\times 10^{-5}}= \frac{1}{\left[1-e_{\rm B,max}^2\left(i_{\rm BC}\right) \right] ^{3/2}}.
\ea
We see this envelope explains very well the observed results from our simulations.

\subsection{Population Synthesis}
\label{sec:pop_syth}

\begin{figure*}[t!]
\center
\includegraphics[width=18cm]{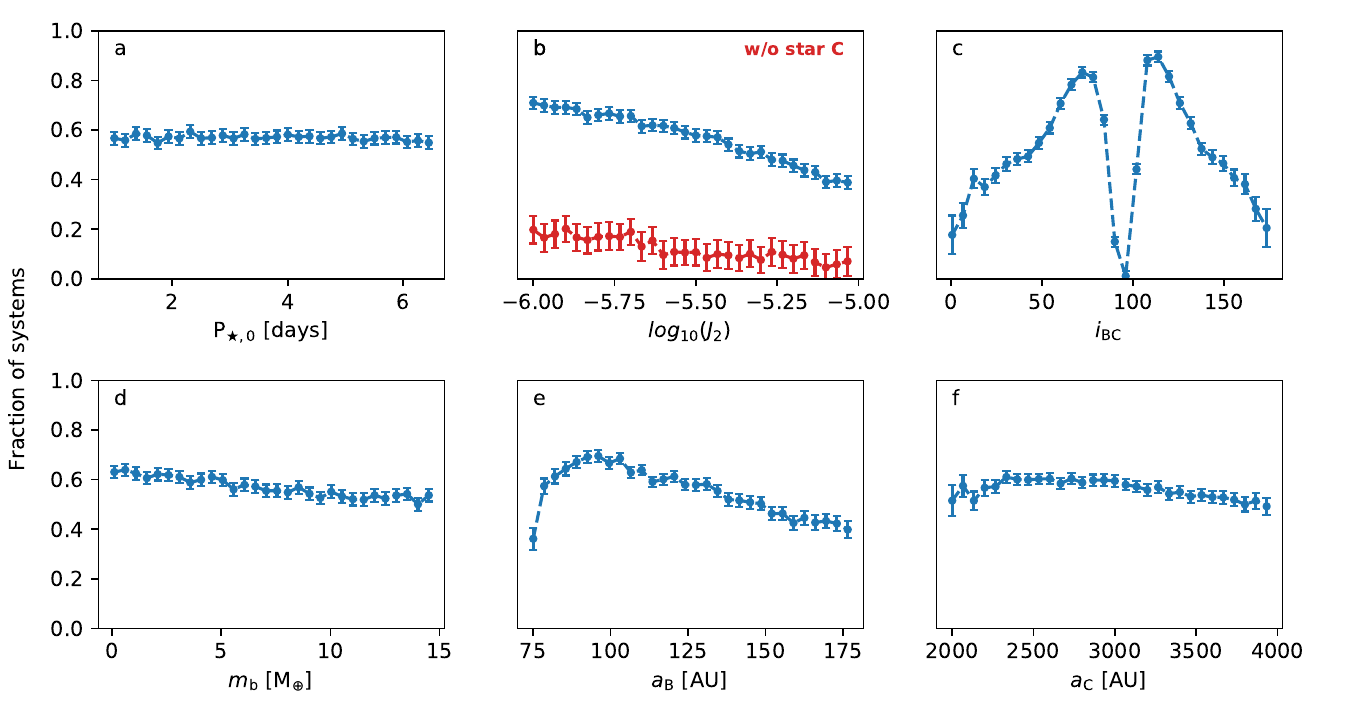}
\caption{Fraction of the systems that reach the observed angle of $124^\circ$ in our population synthesis study described in \S\ref{sec:pop_syth}. Overall, the success rate is $56\%$, while in $17\%$ of the systems the planets are destroyed and $27\%$ never reach $\geq124^\circ$ (if we only consider the systems which are not destroyed, the success rate goes up to 67\%). From \textbf{panels a to f} we show the fraction dependence on: initial period of the star, final value of J2, mutual inclination of B and C, mass of planet b, semi-major axis of stars B and C.
The red marks in panel b correspond to a case without star C for comparison where the success rate is only $12\%$ and the initial distribution of $e_{\rm B}$ (not shown) is largely skewed to very large values with a mean of $0.86$.}
\label{fig:fig4}
\end{figure*}

In Figure \ref{fig:fig4} we explore a parameter space neighbouring the observed values in Figure \ref{fig:schematic} and a range of spin-down and $J_2$ histories (the $J_2$ parameter in Figure \ref{fig:fig4}b refers to the current value for star A, assuming a period of 6.63 days) with $\sim$ 50 000 trials.

As in all previous simulations, the planetary system is initially aligned with the stellar equator ($i_{\rm bc,s}=0$). On the other hand, the stellar orbits are drawn following the following properties:
\begin{itemize}
    \item The inclinations are isotropically oriented $\cos i\sim \mbox{U}(-1,1)$, with uniformly distributed longitude of periapsis and longitudes of the ascending nodes.
    \item As we had no prior knowledge of the current $J_2$ or initial period of the central star, we assumed a log-Uniform and uniform distribution respectively.
    \item The eccentricities of stars B and C were drawn according to the distributions in \cite{Tokovinin}, considering that star C has an inner binary:
\ba
    f(e_{\rm B}) &=& 1.2 e_{\rm B} + 0.4,~ \left<e_{\rm B}\right> = 0.6,\\
    f(e_{\rm C}) &=& 0.24 e_{\rm C} + 0.88,~ \left<e_{\rm C}\right> = 0.52.
\ea
\item The semi-major axes $a_{\rm B}$ and $a_{\rm C}$ were drawn from a log-Uniform distribution, rejecting those that do not satisfy the stability criterion by \citet{Mardling}:
\begin{equation}
\frac{a_{\rm C}(1-e_{\rm C})}{a_{\rm B}}> 2.8\bigg[\left(1 + \frac{M_{\rm C}}{M_{\rm B}+M_{\rm A}}\right)\frac{(1 + e_{\rm C})}{(1 - e_{\rm C})^{1/2}}\bigg]^{2/5}.
\end{equation}

\end{itemize}

Finally, we also reject the systems whose orbit-averaged projected distance
\begin{equation}
D_{\rm proj}=\frac{a(1-e^2)^{5/2}}{2\pi}\int_0^{2\pi}\frac{df\left[1-\sin^2(\omega+f)\sin^2I\right]^{1/2}}{\left[1+e\cos f\right]^3}
\end{equation}
do not fall within the observed $1\sigma$ error bars of $D_{\rm proj, B}= 113 \pm 2$ au and $D_{\rm proj, C}=2467^{+177}_{-155}$ au \citep{Hjorth_2019}.

We present these results in Figure \ref{fig:fig4} as the fraction of total simulations achieving the observed obliquity within the estimated age of the system of $4.0^{+1.6}_{-0.8} $Gyr and find that $56\%$ of the systems could explain K2-290.


Looking into each panel, we see no dependence on initial period (panel a). On panel b we see the expected trend for $J_2$ as for lower values, the planets are less coupled to their star and are more easily knocked over. This trend was already hinted by the white envelope in Figure \ref{fig:fig3}c, where for larger $J_2$ the inclinations needs to be larger to cross the resonances (Eq. \ref{eq:J2_cosI}).

We also observe that the mutual inclination distribution has a dip at i$_{\rm BC}\sim 100^\circ$ (panel c) due to the large eccentricities that disrupt the planetary system, while it reaches two maxima just outside this dip that allow for large $e_{\rm max}$ exploring a wider range of nodal frequencies $\dot{\Omega}_{\rm bc,B}\propto (1-e_{\rm B}^2)^{-3/2}$ to drive chaos.

In turn, panel d shows a slightly higher success rate for lower masses of the inner planet. As this planet represents very little of the angular momentum of the bc system this is only due to the coupling between bc and the central star where both planets contribute on the same order. Finally for panels e and f we see there is a preferred distance for the outer stars. More noticeable for star B.


\section{Discussion} \label{discussion}

In this work, we have studied the inclination history of the multi-planet system orbiting K2-290A perturbed by two distant stars and the rotationally-induced stellar quadrupole. We show that the stellar obliquity increases chaotically, often reaching retrograde configurations and reproducing the observed value of $124\pm6^\circ$.

A striking property of K2-290 is that all nodal precession frequencies for the planets and the star's spin axis are comparable ($\sim0.2-0.3\mbox{ Myr}^{-1}$; equations \ref{eq:Omega_bc_s}-\ref{eq:Omega_bc_B}). Thus, secular resonances are often crossed, triggering secular chaos, a process that is assisted by the ZKL oscillations of the AB binary driven by star C.

Other results include:
\begin{itemize}
    \item The parameter space where retrograde obliquities are attained grows by including the effect of the distant star C. Without C we cannot reproduce the observed obliquity for our estimated current value of $J_2$ (Panel b of Figure \ref{fig:fig3}), unless $e_{\rm B}$ is initially very large. A population synthesis without star C shows a success rate of only $12\%$ (compared to $56\%$ with C) and the successful systems have $\left <e_{\rm B}\right>\simeq0.86$ (Figure \ref{fig:fig4}).
    \item We show that a necessary condition for secular chaos is that $i_{\rm BC}$ is sufficiently large to drive larger values of $e_{\rm B, max}$ (see equation \ref{eq:J2_cosI} and envelope in panel c of Figure \ref{fig:fig3}). As such, we expect that B and C are highly inclined relative to each other (see panel c in Figure \ref{fig:fig4}).

    \item We see some dependence on the final value of $J_2$ (panel b, in Figure \ref{fig:fig4}), but no dependence on initial period of the star (panel a). In other words, the mechanism is independent on the initial value of $J_2$. As for the current value, the somewhat evolved star with radius of $1.51R_\odot$ would have had a radius closer to $1.25R_\odot$ during its main sequence; a factor of $\sim1.8$ lower in the value of $J_2$.
\end{itemize}

In what follows, we discuss how our main findings depend on the presence of another planet and how they fit in the bigger picture of obliquity excitation.

\subsection{Presence of other planets in K2-290}

Even though only two planets (K2-290b and K2-290c) have been discovered around K2-290A, there is still the possibility that more planets could be discovered in the future.

Using RV data from HARPS-N, \citet{Hjorth_2021} find a radial acceleration for star A of $\dot{\gamma}=9\pm5\mbox{ m} \mbox{ s}^{-1} \mbox{ yr}^{-1}$. This acceleration, although compatible with zero within 2$\sigma$ and with the presence of star B (which would cause an acceleration on the order of $5\mbox{ m} \mbox{ s}^{-1} \mbox{ yr}^{-1}$) within 1$\sigma$, could also be caused by other planets orbiting K2-290A with a longer period.

The main effect of this undiscovered planets would be to increase the coupling between the planetary system and star B (increase $\dot{\Omega}_{\rm bc,B}$). Thus, in order to encounter the resonance $\dot{\Omega}_{\rm s,bc}\sim\dot{\Omega}_{\rm bc,B}$, it would demand a larger value of $J_2$, possibly earlier in the evolution of the star.

\subsection{Link to retrograde hot Jupiters}

The chaotic spin-orbit behavior observed in this work is reminiscent of the work by \citet{storch2014,storch2015} in the context of Jovians undergoing ZKL cycles and high-eccentricity migration to hot Jupiter orbits.
In their work, the Jovian itself plays the role of both the star B undergoing eccentricity oscillations and the planetary system that torques the star.

A qualitative difference occurs when the planet(s) do not have enough angular momentum to torque the star ($|\dot{\Omega}_{\rm s,bc}|\ll |\dot{\Omega}_{\rm bc, s}|$). If so, the set-up in \citet{storch2014} with a migrating  planet will not drive chaotic obliquity evolution, but just regular oscillations due to the ZKL cycles. Instead, in our model the periodic resonance crossings for the planetary orbits $\dot{\Omega}_{\rm bc, s}\sim\dot{\Omega}_{\rm bc, B}$ may occur at all planetary masses (Eq. \ref{eq:Omega_bc_s} and \ref{eq:Omega_bc_B}). In other words, our model may lead to either chaotic tumbling of the star as in \citet{storch2014} and/or the planetary orbits, both leading to chaotic obliquity evolution. We have checked this by decreasing the masses of our planets by a factor of 10 or 100.


\subsection{Primordial disk misalignment or tertiary-driven secular chaos?}

As discussed by \citet{Hjorth_2021}, K2-290 is the first system that provides strong support to the primordial disk misalignment theory \citep{batygin12,lai2014,Spalding2014}. The companion star B lies at a desired distance of $\sim 100$ au; not too close to suppress planet formation \citep{MK2019} and not too far to require unrealistically long disk lifetimes.

Unlike the primordial tilting where a resonance crossing is mediated by the dispersing disk, our proposal to cross the resonances relies on the wider-orbit companion star C. This is an important difference as the star C is currently observed and large obliquities are excited by a wide range of orbital distances ($\sim 2000-4000$ au, Figure \ref{fig:fig4}f) that are consistent with the observations. The chaotic obliquity excitation is then a natural outcome of the observed system, relying just on few-body gravitational dynamics and with a success rate of $\sim40-70\%$ depending on the assumed parameters, mainly the value of $J_2$. This is a difficult value to estimate as it depends on the internal structure of the star \citep{batygin2013}, but we still observe the effect for a wide range of values. As for the disk, involving gas dynamics and evaporation processes, the dynamics is far more uncertain.

Finally, we remark that the primordial tilting and our proposed model can work in tandem. The chaotic obliquity excitation described here starts taking place long after the disk disperses and it is fairly insensitive to the initial stellar obliquity.

\subsection{Future testable predictions}

Figure \ref{fig:fig4}c shows this mechanism is more effective when the orbits of stars B and C are almost perpendicular to each other. Although this mutual inclination decreases when $e_{\rm B}$ grows, the system spends more time with its large-inclination state, as can be seen in Figure \ref{fig:fig1}c. Thus, a prediction of our results would be that $i_{\rm BC}$ should be large, peaking at $\sim 80^\circ$ or $\sim 110^\circ$.

As shown by \cite{triple}, the orbital planes are uncorrelated for tertiary stars at $\sim10^3$ au (Figure 1 therein),
corroborated for larger values of the outer separation (up to $10^4$ au) in \citet{tokovinin21}.


Given a Keplerian speed for star C (relative to A) of $\sim$0.7 km/s, the relative RVs can be measured with current spectrographs. Also we estimate a proper motion of $\sim$0.6 mas/yr for motion in the sky plane which is above Gaia's uncertainties.


\subsection{Applications to other planetary systems}

KOI-5 \citep{koi5} is a more compact triple star system with a confirmed inclined inner planet (and an unconfirmed outer one). We ran a few simulations to test if the effects previously mentioned could also explain this system and found that with a sufficiently large mutual inclination of the external star, we can even get retrograde states for KOI-5's planets.

It is worth noting that there is nothing special about the measured stellar obliquity of 124 degrees in our simulations. In fact, given this mechanism, we could expect to find any retrograde angle up to 180 degrees in other systems with a similar architecture.

More generally,  the dynamics depend on the outer bodies only through the amplitude of their tidal fields $\propto M/a^3$. Therefore, we may replace the star B in our set-up for a Jovian-mass planet $\sim10$ times closer to get the same behavior. Correspondingly, making sure star C is close enough to drive ZKL oscillations of such planet (i.e., unquenched by the inner planets).

\section{Conclusions} \label{conclusions}

In this work we show that the striking retrograde stellar obliquity of planets K2-290b and c (124$\pm$6 degrees) is a natural outcome of the long-term spin-orbit dynamics driven by the distant stars B and C. Here, the star C drives eccentricity and/or inclination oscillations in the AB orbit, triggering widespread chaos in the evolution of the planets inclinations and stellar obliquities.

Since our model works for a wide range of initial conditions, independent of the host star's spin history, we suggest that the previously proposed explanation relying on the primordial tilt of the birth protoplanetary disk may not be required.


Finally, we remark that, for our estimate of J$_2$ for the central star, we can only achieve the observed obliquity if star C is present, and so, the effect of this perturber, even though it is very distant to the planetary system, should not be ignored.


\acknowledgements

We thank the anonymous reviewer for their feedback, which
improved this manuscript. The authors would also like to thank Julio Chanamé, Gijs Mulders,  Diego Muñoz, Smadar Naoz, Barbara Rojas Ayala and J. J. Zanazzi for their helpful comments.
C.P. acknowledges support from ANIDMillennium Science Initiative-ICN12\_009, CATA-Basal AFB-170002, ANID BASAL project FB210003, FONDECYT Regular grant 1210425, and CONICYT+PAI (Convocatoria Nacional subvencion a la instalacion en la Academia convocatoria 2020, PAI77200076). S.B. acknowledges additional support from the National Agency for Research and Development (ANID) / Scholarship Program / Doctorado Nacional grant 2021 - 21211921.

\bibliography{refs}

\appendix

\section{Equations of motion and definitions}\label{app:equations_of_motion}

To describe the orbits of each body we use the vectors of eccentricity ${\bf e}=e{\bf \hat{e}}$ and specific angular momentum ${\bf j}=(1-e^2)^{1/2}{\bf \jhat}$ with the convention of ${\bf e}$ pointing in the direction of the periapsis and ${\bf j}$ pointing in the direction of the angular momentum. Then, given the potential in terms of these vectors (from \citealt{TY2014}), we can get their secular evolution:
\ba
\label{eq:full_hamiltonian}
\phi =
-\frac{\phi_{\rm bc,A}}{2} \left[
({\bf \hat{s}}\cdot \jin)^2-\tfrac{1}{3}
\right]
-\frac{\phi_{\rm bc,B}}{2} \left[({\bf j}_{\rm bc} \cdot \jhat_{\rm B} )^2 -\tfrac{1}{3}\right]
-\frac{\phi_{\rm B,C}}{2} \left[-5( {\bf e}_{\rm B} \cdot \jhat_{\rm C} )^2 +({\bf j}_{\rm B} \cdot \jhat_{\rm C} )^2
+2e_{\rm B}^2-\tfrac{1}{3}\right]
\ea

where the amplitudes are given by:

\ba
\phi_{\rm bc,A}&=&\frac{3J_2G M _{\rm A} R _{\rm A}^2}{2} \left( \frac{M_{\rm b}}{a_{\rm b}^3} + \frac{M_{\rm c}}{a_{\rm c}^3} \right),\\
\phi_{\rm bc, B}&=&\frac{3G M_{\rm B} M_{\rm c} a_{\rm c}^2}{4b^3_{\rm B}},\\
\phi_{\rm B, C}&=&\frac{3G M_{\rm A} M_{\rm B} M_{\rm C} a_{\rm B}^2}{4(M_{\rm A}+M_{\rm B})b^3_{\rm C}},
\ea
and $b=a(1-e^2)^{1/2}$ is the semi-minor axis.

We solve the motion using the Milankovitch set of equations (e.g., \citealt{TY2014}) as
\ba
\label{eq:Milanko_1}
\frac{d{\bf s}}{dt}&=& \frac{1}{S}\left(\nabla_{{\bf s} } \phi \times {\bf s} \right)
\\
\frac{d{\bf j}_{\rm k}}{dt}&=& \frac{1}{L_{\rm k}}\left(\nabla_{{\bf j}_{\rm k} } \phi \times {\bf j}_{\rm k} +
\nabla_{{\bf e}_{\rm k} } \phi \times {\bf e}_{\rm k}\right),
\\
\frac{d{\bf e}_{\rm k}}{dt}&=& \frac{1}{L_{\rm k}}\left(\nabla_{{\bf e}_{\rm k} } \phi \times {\bf j}_{\rm k} +
\nabla_{{\bf j}_{\rm k} }\phi \times {\bf e}_{\rm k}\right),
\ea
where the sub-index $\mbox{k}=\left\{ {\rm bc, B, C} \right\}$ \footnote{Except for the eccentricity of the planets. As the potential does not depend on this parameter (which starts at zero), the planetary orbits remain circular, as was required by the fact that both planets are coupled.} and
$L_{\rm bc}=M_{\rm b}\sqrt{GM _{\rm A} a_{\rm b}} + M_{\rm c}\sqrt{GM _{\rm A} a_{\rm c}} $, $L_{\rm B}=\frac{M_{\rm B}M_{\rm A}}{M_{\rm B}+M_{\rm A}}\sqrt{GM_{\rm A} a_{\rm B}}$ and $L_{\rm C}=\frac{M_{\rm C}(M_{\rm A}+M_{\rm B})}{M_{\rm C}+M_{\rm B}+M_{\rm A}}\sqrt{G (M_{\rm A}+M_{\rm B}) a_{\rm C}}$ are the orbital angular momenta for circular orbits and $S = k_q M_{\rm A} R_{\rm A}^2 \Omega_{\star}$ is the magnitude of the angular momentum spin of the central star.

\end{document}